# Coherence analysis of phase-controlled HOM effects

Byoung S. Ham

Department of Electrical Engineering and Computer Science, Gwangju Institute of Science and Technology
123 Chumdangwagi-ro, Buk-gu, Gwangju 61005, South Korea
(Submitted on February 19, 2025; bham@gist.ac.kr)

**Abstract**
The second-order intensity correlation of entangled photons has been intensively studied for decades, particularly for the Hong-Ou-Mandel (HOM) effect and nonlocal correlation - key quantum phenomena that have no classical counterparts. Recently, a path-entangled two-photon state has been experimentally demonstrated for both bosonic (symmetric) and fermionic (anti-symmetric) HOM effects by manipulating the photon's phase at one input port. Entanglement represents a quantum superposition of path- or energy-correlated two-photon states with a relative phase. According to the conventional quantum mechanics, this phase is not an individual property but collective attribute of interacting photons. Here, the wave nature of photons is employed to coherently analyze the phase-controlled HOM effects recently observed in *npj Quantum Info*. **5**, 43 (2019). A pure coherence approach is applied to derive a general solution for these phase-controlled HOM effects. Consequently, the quantum mystery of HOM effects, traditionally interpreted through the particle nature of quantum mechanics, is revealed as a coherent phenomenon between entangled photons via a selective choice of correlated photons.

1. ## Introduction

Over the past decade, quantum information science [1-3] has been successfully applied to quantum technologies such as quantum computing [4,5] and quantum communications [6-8]. At its foundation, quantum information relies on quantum superposition [9] and quantum entanglement [10], where superposition pertains to a single particle [11] and entanglement involves multiple particles [2]. More importantly, quantum superposition underpins quantum entanglement, as seen in systems like the double-slit experiment or a beam splitter (BS). As Feynman stated, most mysterious phenomenon in quantum mechanics is a single photon's self-interference, since a photon cannot be split into two [12]. Consequently, quantum superposition is understood as the probability amplitudes distributed across both paths [11]. This concept has led to the quantum eraser, which addresses the which-way information of a single photon in an interferometer [13,14]. In quantum mechanics, the energy-time uncertainty relation is analogous to the number-phase uncertainty relation. The origin of the number-phase relation lies in de Broglie's wave-particle duality, where wave and particle natures are mutually exclusive [15]. As a result, conventional quantum information science, which is based on the particle nature, has traditionally overlooked the role of wave nature [11].

The Hong-Ou-Mandel (HOM) effect is one of the most fundamental phenomena in two-photon intensity correlation in a BS [16,17]. This two-photon interference has been a major research topic in quantum information science due to the weird quantum phenomenon of photon bunching, where photons coalesce into the same output port, which is known to be impossible by any classical means [18,19]. While the HOM effect has been extensively studied using quantum operators [11], the BS matrix formalism originates from pure coherence optics [20]. According to wave-particle duality, quantum operators inherently lose the phase information of coupled photons, even though preserving the phase properties of the BS remains essential [11]. This highlights a fundamental limitation of the conventional quantum approach to the HOM effect – namely, the absence of phase relationships between paired photons. However, in reality, photon pairs generated via spontaneous parametric down-conversion (SPDC) arise from coherence optics, governed by the phase-matching conditions among pump, signal, and idler photons [21-23].

Recently, new approaches to interpreting the HOM effect have emerged, using coherent photons [24-28]. Within the framework of wave-particle duality in quantum mechanics, the choice of the wave nature is optional [27,28]. Schrödinger's wave equation, inspired by de Broglie's wave-particle duality, provides a wave representation of a single particle [15]. Mathematically, a single particle can be described as a wave packet formed by the superposition of multiple monochromatic waves, governed by Maxwell's equations. The bandwidths of the superposed monochromatic waves determines the temporal and spatial localizations of the wave packet, which is conceptually equivalent to the probability-amplitude of a single particle in quantum mechanics [11]. This suggests that the phase nature of a single photon does not need to be disregarded, in spite of its long-standing neglect in quantum information science. Here, the phase nature of quantum mechanics is applied to the recently observed phase-controlled HOM effects [19]. Through coherence analysis, the so-called quantum mystery is revealed as a phase relationship between entangled photons. Moreover, the HOM dip and HOM peak are directly understood as consequences of phase manipulation between photon pairs. This coherence-based interpretation of the HOM effects offers a clearer and deeper understanding of quantum mechanics.

2. Results

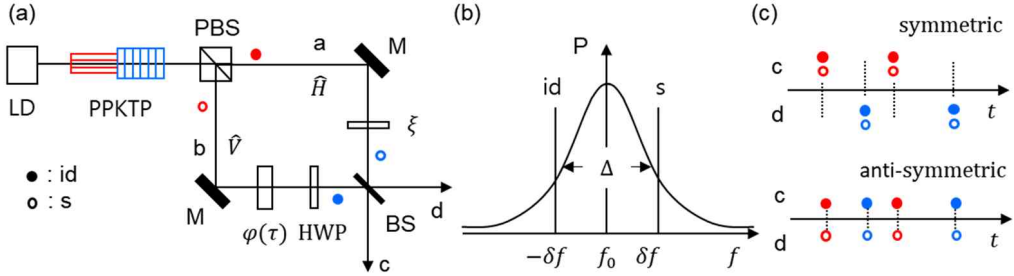

**Fig. 1.** Schematic of two-photon interference on a beam splitter. (a) $\xi$ phase-controlled HOM setup. (b) SPDC-generated photon spectrum. (c) Mode selective entangled photon pairs. BS: 50/50 nonpolarizing beam splitter, HWP: 45°-rotated half-wave plate. LD: laser diode, PBS: polarizing BS. s: signal, id: idler. Δ: bandwidth.

Figure 1(a) is a schematic of a modified two-photon interference setup for $\xi$-controlled HOM effects, where $\xi$ represents the phase parameter between entangled photons [18]. An entangled pair of orthogonally polarized signal (s) and idler (id) photons is generated via a type-II SPDC process using two periodically poled potassium titanyl phosphate (PPKTP) crystals [18]. The second PPKTP crystal is rotated by 90° to the first one. Although the signal and idler photons are initially orthogonally polarized and spatially separated by a polarizing BS (PBS), each path of Mach-Zehnder interferometer (MZI) contains both signal and idler photons in the same polarization basis due to the 90°-rotated second PPKTP crystal [18]. Consequently, both inputs of BS maintain a uniform distribution of signal and idler photons. By a single pump photon whose frequency is $2f_0$, each pair of photons denoted by red and blue dots is generated at an equal probability. A half-wave plate (HWP) rotated at 45° converts originally vertically polarized photons into horizontally polarized ones, making them indistinguishable on BS. This results in formation of a path-entangled two-photon state $|\psi\rangle$, given by $|\psi\rangle = (|s\rangle_a|id\rangle_b + e^\xi |id\rangle_a|s\rangle_b)/\sqrt{2}$, where $|s\rangle_a|id\rangle_b$ and $|id\rangle_a|s\rangle_b$ represent path-correlated signal-idler photon pairs impinging on BS through input paths 'a' and 'b.' The plus sign in $|\psi\rangle$ signifies quantum superposition between these correlated photon pairs, while the phase term $e^\xi$ serves as a coherent controller between the superposed terms. By adjusting $\xi$, the HOM measurement can yield either a symmetric (bosonic-like) HOM dip or an anti-symmetric (fermionic-like) HOM peak [18]. Figure 1(c) illustrates the $\xi$-dependent photon bunching (top) and anti-bunching (bottom) effects at the BS output ports (see analysis). At a low generation rate of path-entangled photon pairs - ensuring incoherence between pairs - no first-order interference fringe appears at either output port of MZI, resulting in local randomness (see analysis).

Here, Fig. 1(a) is coherently analyzed to examine the path-entangled two-photon state $|\psi\rangle$ and to reveal the underlying quantum nature of the HOM effects. To achieve this, a relative phase $\zeta$ between entangled photons is introduced, which remains consistent with quantum mechanics, even though it would violate the number-phase uncertainty relation for a single photon.

For the coherence analysis, the path-entangled photon pair $|\psi\rangle$ is treated as a monochromatic wave, similar to a Gaussian-distributed Fourier series, where the bandwidth determines a spatial and temporal localization of the photon wave packet. These Gaussian-distributed photon pairs are randomly generated by the SPDC process, forming a statistical ensemble. According to the energy and momentum conservation laws in second-order nonlinear optics, the polarization-correlated signal and idler photons in Fig. 1(a) are temporally and spatially overlapped in a degenerate scheme [18]. However, the first and second correlation terms in $|\psi\rangle$ cannot coexist simultaneously by definition and coincidence measurements. Therefore, in this coherence analysis, these two terms are considered coherently but separately. For simplicity, a specific jth photon pair from the Gaussian distribution of Fig. 1(b) is examined, where the pair exhibits spectral symmetry with $\pm \delta f_j$ across $f_0$, satisfying $2f_0 = f_s + f_{id}$. A delay time $\tau$ is introduced in path 'b,' resulting in a phase shift $\varphi$, where $\varphi_j = \delta f_j \tau$. Due to the random $\delta f_j$, the phase $\varphi_j$ varies statistically across photon pairs unless $\tau = 0$. Furthermore, from a measurement perspective, the interference of photon pairs is dominated by a beating phenomenon in the coincidence detection, where $\varphi$ depends on the bandwidth $\Delta$ (see analysis). The parameter $\xi$, applied to the second term in $|\psi\rangle$, is assigned to either the signal or idler photon in all pairs. Thus, $\xi$ can be set in any path of MZI. In Fig. 1, path 'a' is chosen for the analysis, as experimentally demonstrated in ref. 18.

*Analysis*

For the jth SPDC-generated entangled photon pairs, the following coherence relation can be established using the BS matrix [20] for $|s\rangle_a |id\rangle_b$ in $|\psi\rangle$:

$$\begin{bmatrix} E_c \\ E_d \end{bmatrix}_j = \frac{1}{\sqrt{2}} \begin{bmatrix} 1 & i \\ i & 1 \end{bmatrix} \begin{bmatrix} E_a \\ E_b \end{bmatrix}_j$$
$$= \frac{E_0}{\sqrt{2}} e^{i(kx - 2\pi f_0 t)} e^{i\Delta_j} \begin{bmatrix} 1 & i \\ i & 1 \end{bmatrix} \begin{bmatrix} e^{i\xi_j} \\ e^{i(\zeta - 2\Delta_j)} \end{bmatrix}, \quad (1)$$

where $\Delta_j = \delta f_j \tau = \varphi_j$ is the controlled delay time $\tau$-induced phase in path 'b' [18]. With the wave nature, the conventional quantum operator is replaced by the electromagnetic field $E_k$, where $E_0$ represents the amplitude of a single photon. A relative phase $\zeta$ between entangled photons is introduced and assigned to the idler photon. In Eq. (1), the signal and idler photons are correlated with input paths 'a' and 'b,' respectively.

For the analysis of the second correlation term in $|\psi\rangle$, the input paths are simply swapped, which corresponds to the matrix element exchange in the final term of Eq. (1):

$$\begin{bmatrix} E_{c'} \\ E_{d'} \end{bmatrix}_j = \frac{1}{\sqrt{2}} \begin{bmatrix} 1 & i \\ i & 1 \end{bmatrix} \begin{bmatrix} E_b \\ E_a \end{bmatrix}_j$$
$$= \frac{E_0}{\sqrt{2}} e^{i(kx - 2\pi f_0 t)} e^{i\Delta_j} \begin{bmatrix} 1 & i \\ i & 1 \end{bmatrix} \begin{bmatrix} e^{i(\zeta - \xi_j - 2\Delta_j)} \\ 1 \end{bmatrix}, \quad (2)$$

where the minus sign in $\xi_j$ arises from $-\delta f_j$ of the idler photon.

From Eqs. (1) and (2), the following relationships are derived:

$$E_c^j(\tau) = \frac{E_0}{\sqrt{2}} e^{i(kx - 2\pi f_0 t + \Delta_j)} e^{i\xi_j} \left(1 + i e^{i(\zeta - \xi - 2\Delta_j)}\right), \quad (3)$$

$$E_d^j(\tau) = \frac{iE_0}{\sqrt{2}} e^{i(kx - 2\pi f_0 t + \Delta_j)} e^{i\xi_j} \left(1 - i e^{i(\zeta - \xi - 2\Delta_j)}\right). \quad (4)$$

$$E_{c'}^j(\tau) = \frac{E_0}{\sqrt{2}} e^{i(kx - 2\pi f_0 t + \Delta_j)} e^{i\xi_j} \left(e^{i(\zeta - \xi - 2\Delta_j)} + i\right), \quad (5)$$

$$E_{d'}^j(\tau) = \frac{iE_0}{\sqrt{2}} e^{i(kx - 2\pi f_0 t + \Delta_j)} e^{i\xi_j} \left(e^{i(\zeta - \xi - 2\Delta_j)} - i\right). \quad (6)$$

Thus, the corresponding jth output intensities for the jth photon pair are as follows:

$$I_c^j(\tau) = I_0 \left(1 - \sin(\zeta - \xi - 2\Delta_j)\right), \quad (7)$$

$$I_d^j(\tau) = I_0 \left(1 + \sin(\zeta - \xi - 2\Delta_j)\right). \quad (8)$$

$$I_{c'}^j(\tau) = I_0 \left(1 + \sin(\zeta - \xi - 2\Delta_j)\right), \quad (9)$$

$$I_{d'}^j(\tau) = I_0 \left(1 - \sin(\zeta - \xi - 2\Delta_j)\right). \quad (10)$$

For $\xi = 0$, $I_c^j(\tau) = I_0(1 - \sin(\zeta - 2\Delta_j)) = I_{d'}^j(\tau)$ and $I_d^j(\tau) = I_0(1 + \sin(\zeta - 2\Delta_j)) = I_{c'}^j(\tau)$ are obtained. For $\xi = \pi/2$, the intensities are $I_c^j(\tau) = I_0(1 + \cos(\zeta - 2\Delta_j)) = I_{d'}^j(\tau)$ and $I_d^j(\tau) = I_0(1 - \cos(\zeta - 2\Delta_j)) = I_{c'}^j(\tau)$. Thus, the uniform intensity in each output port of BS is coherently satisfied for all N entangled photon pairs, based on the sum intensities $I_k^j(\tau) + I_{k'}^j(\tau)$:

$$\langle I_c(\tau) \rangle = \frac{1}{2N}\sum_{j=1}^{N}[I_c^j(\tau) + I_{c\prime}^j(\tau)] = I_0, \quad (11)$$

$$\langle I_d(\tau) \rangle = \frac{1}{2N}\sum_{j=1}^{N}[I_d^j(\tau) + I_{d\prime}^j(\tau)] = I_0. \quad (12)$$

Due to the self-cancellation of the sine and cosine terms, Eqs. (11) and (12) demonstrate the local intensities induced by basis randomness in the HOM effect [16-19]. This confirms the validity of the current coherence approach. If only one term in $|\psi\rangle$ is selectively chosen in Fig. 1, e.g., using one PPKTP, however, the local randomness with uniform intensity should be violated, as shown in Eqs. (7)-(10), even though no change results in the HOM effect (see below)

For the second-order intensity correlation $R_{cd}^j(\tau)$ and $R_{c\prime d\prime}^j(\tau)$ through coincident detection of output photons in Fig. 1(a), the HOM effect is related to the average of coincidence measurements across all statistically independent photon pairs. Unlike uniform intensity $I_0$ in each output port of BS for entangled photon pairs in $|\psi\rangle$, the $\tau$-dependent second-order intensity correlation is given by the following for $\xi = 0$ (see the blue curve in the left panel of Fig. 2):

$$\langle R_{cd}(\tau)\rangle = \langle R_{c\prime d\prime}(\tau)\rangle = \frac{1}{N}\sum_{j=1}^{N} I_c^j(0)I_d^j(\tau)$$
$$= \frac{I_0^2}{N}\sum_j^N \cos^2(\zeta - 2\delta f_j \tau). \quad (13)$$

At $\tau = 0$, $\zeta = \pm\pi/2$ must be satisfied in Eq. (13) for the symmetric (bosonic) HOM dip [18,19]. Here, $\Delta_j$ is affected by $\tau$ due to the term $\delta f_j \tau$. The common optical harmonic oscillation term has already been factored out in the first-order intensity correlations, so the difference frequency factor becomes the dominant term in Eqs. (7)-(10). The HOM dip represents a photon bunching phenomenon, where photons bunch into either output port of BS, as shown in the upper panel of Fig. 1(c) [18,19,25]. Thus, the symmetric HOM dip arises from the inherent $\pi/2$ phase shift between the signal and idler photons for all pairs, with the selective measurement through coincidence detection. Here, $\langle R_{cd}(\tau)\rangle = 1/2$ in Eq. (13) is the direct consequency of the incoherence feature at $\tau \gg \Delta^{-1}$.

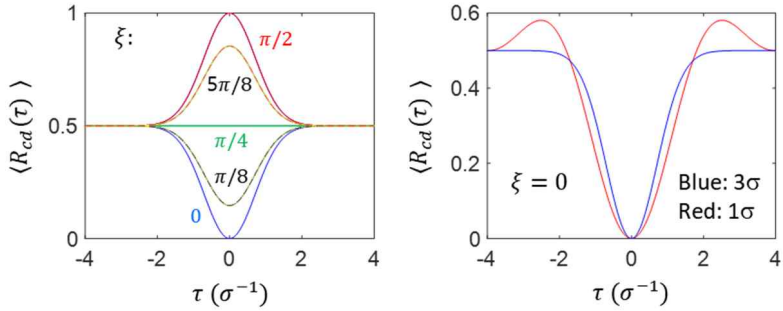

**Fig. 2.** Numerical calculations of $\xi$-controlled HOM effects. Photons are Gaussian distributed: $\sigma = 10^9$. Photons are Gaussian distributed and covered by $\pm 3\sigma$ for calculations in the left panel.

For $\xi = \pi/2$, the $\tau$-dependent second-order intensity correlation is given by (see the red curve in the left panel of Fig. 2):

$$\langle R_{cd}(\tau)\rangle = \langle R_{c\prime d\prime}(\tau)\rangle = \frac{1}{N}\sum_{j=1}^{N} I_c^j(0)I_d^j(\tau)$$
$$= \frac{I_0^2}{N}\sum_j^N \sin^2(\zeta - 2\delta f_j \tau). \quad (14)$$

Under the same condition of $\tau = 0$ and $\zeta = \pm\pi/2$, Eq. (14) results in the anti-symmetric HOM peak [18,19,25]. From Eqs. (13) and (14) not only the relative phase $\zeta$ between entangled photons, but also the role of $\xi$ is coherently confirmed for both symmetric and anti-symmetric HOM effects, which have already been experimentally demonstrated [18,19,25]. As a result, the phase nature of photons unveils otherwise the mysterious HOM effects under the particle nature, where the fixed phase relation between path-entangled photons serves as the origin of the HOM effect. Furthermore, controlling the phase ($\xi$) of one photon in an entangled pair for one input port of BS in Fig. 1(a) allows for coherent conversion of the HOM dip into the HOM peak. Since this phase is coherently controllable, the HOM dip and peak represent two extremes of all $\xi$-dependent HOM effects, as shown in the left panel of Fig. 2 [29]. The oscillation of the HOM effect [17] arises when partial events in the spectral domain of Fig. 1(b) are selected, as shown in the right panel of Fig. 2. Without phase nature of the entangled pair, the analytically confirmed HOM effects would remain unclear. These individual $\xi$-based notations in Eqs. (13) and (14) are represented by a cumulative factor of $2\xi$ in the Bell states of path-entangled two-photon state $|\psi\rangle$, where $|\psi\rangle = (|s\rangle_a|d\rangle_b + e^{i2\xi}|d\rangle_a|s\rangle_b)/\sqrt{2}$. Similarly, the intensity correlation between MZI outputs in quantum sensing with N00N states is represented in the same way as $R^{(N)} = (1 + \cos N\varphi)/2$, where N is number (order) of entangled photons (intensity correlation) [30].

## 3. Discussion

*Local randomness:*

Regarding the coherence approach for the HOM effects, quantum superposition of a single photon's self-interference plays an essential role. When the signal and idler photons are coincident on BS with $\pm\zeta$ phase shift, the output photons result from the superposition of these two phase coherent interference cases. The '$\pm$' sign in $\zeta$ arises from the swapping of the signal and idler photons due to the PPKTP crystal setup and the PBS in Fig. 1(a), where each input port of BS contains both signal and idler photons with equal probability. Since the global phase of the entangled photon pairs remains unaffected, $\zeta$ plays a critical role in the HOM effects. Through the path exchange of entangled photons provided by the PPKTP setup via PBS, the relative phase between the two input photons on BS must satisfy $\pm\zeta$, where $\zeta$ represents the relative phase of the idler photon with respect to the signal photon. With $\zeta = \pi/2$, as derived in Eq. (13), local randomness is ensured due to the out-of-phase relation between first-order intensity correlations for $\zeta = \pm\pi/2$, as shown in Eqs. (7) and (9). In other words, summing the two opposite cases with $\zeta = \pm\pi/2$ eliminates individual fringes for all pairs. If the signal and idler photons are not mixed in the input ports of BS, local randomness is prohibited, even though the HOM effect persists. Since path-photon entanglement governs photon exchange between two input paths of BS, entangled photons must satisfy the local randomness.

*No wavelength-dependent two-photon interference:*

The HOM effect also originates from $\zeta = \pm\pi/2$, where the intrinsic $\pi/2$ phase difference between the transmitted and reflected photons at BS is added to $\zeta$, resulting in phase shifts of $\pi$ and 0 for the reflected photon. This out-of-phase relation within the same path leads to local randomness, as discussed earlier. On the other hand, the transmitted photon at BS experiences phase shifts of 0 and $\pi$, respectively. Consequently, for $\zeta = \pm\pi/2$, photon bunching consistently occurs at either output port of BS. Thus, the second-order intensity correlation between the two output photons of BS produces the HOM dip. Coincidence detection serves as a selective measurement to distinguish between the two cases of $\zeta = \pm\pi/2$, both of which exhibit the same HOM dip. If one input port of BS is phase-controlled by $\xi$, the phase relation of the output photons also depends on $\xi$. As analyzed in the text, when $\xi = \pi/2$ is met in the upper input port of BS, the reflected photon acquires phase shifts of $3\pi/2$ and $\pi/2$ for $\zeta = \pm\pi/2$. Likewise, the transmitted photon experiences phase shifts of $\pi/2$ and $3\pi/2$, respectively. In both cases, the probability of finding photons in either output port of BS remains equal at 50%, leading to the HOM peak in the intensity product. Thus, the HOM dip and HOM peak are direct consequences of the wave nature, i.e., coherence optics. The absence of wavelength-dependent fringes in the HOM effects stems from the beating phenomenon, whereas spectrally filtered photons contribute to beat frequency-based oscillations, as observed in many HOM experiments.

## 4. Conclusion

The phase-controlled HOM effects were coherently analyzed for the observed symmetric HOM dip and anti-symmetric HOM peak phenomena observed in ref. 18. Unlike the conventional phase-independent interpretation based on the particle nature, the critical condition of the relative phase between entangled photons was coherently derived using the wave nature of photons compatible to the BS matrix for the observed HOM effects. The anti-symmetric HOM peak was also analytically confirmed within the coherence framework, demonstrating that phase control of either photon in the entangled pair was essential in determining the HOM effect. In other words, the HOM dip and HOM peak directly resulted from the phase (ξ) control between entangled photons. To work with this phase control, the assignment of the phase relation (ζ) between entangled photons was a prerequisite. Based on this, both the HOM dip and HOM peak were identified as two extremes of ξ-transient two-photon interference. From the analytically derived coherence solutions, the origin of local randomness in HOM effects was also identified in the random bases of signal and idler photons, leading to an out-of-phase fringe relationship at the BS output ports for the first-order intensity correlations. In spite to the local randomness, the two-photon intensity correlation revealed the ξ-dependent coherence feature of the HOM effects through selective choice of correlated photon pairs in $|\psi\rangle$ via coincidence detection. The absence of wavelength-dependent fringes in the HOM effect was attributed to frequency cancellation between entangled photon pairs, wherein SPDC bandwidth-dependent optical beating dominated. Thus, the coherence approach unveiled the quantum mystery behind the HOM effects, where the relative phase between entangled photons played a crucial role.

**Declarations**
Ethical Approval and Consent to participate: Not applicable.

**Availability of data and materials**
The data that support the findings of this study are available upon reasonable request from the authors.
**Competing interests**
The author declares no competing interests.

**Funding:** This work was supported by the IITP-ITRC (IITP-2025-RS2021-II211810) funded by the Korea government (Ministry of Science and ICT).

**Author's contribution:** BSH conceived the idea and wrote the paper.